\documentclass{ws-procs975x65}

\begin{document}

\title{Relativistic Mean Field Models at High Densities}

\author{A. Sulaksono, P.T.P. Hutauruk, C.K. Williams, and T. Mart}

\address{Departemen Fisika, FMIPA, Universitas Indonesia,
Depok 16424, Indonesia}

\maketitle

\abstracts{The effects of vector-isovector terms predicted by relativistic mean field models based on effective field theory  on the equation of state and the neutrino mean free path in neutron stars have been studied. Using a procedure similar to Ref.\,\cite{pieka}  in treating the isovector sector, a  parameter set (G2*) that predicts a much smaller proton fraction than the standard one (G2) can be obtained. We found that the G2* has a softer equation of state compared to the G2 at higher densities and the disappearance of the anomalous behavior of the neutrino mean free path in neutron star does not depend on the proton fraction.}

\section{Introduction}
The relativistic mean field (RMF) model allows ground state properties of finite nuclei to be described with relatively high precision (for a review see, e.g., Ref.\,\cite{pg}). If we extrapolate this model to the high density regime, it is known that the RMF model predicts a much stiffer equation of state (EOS) compared to the Bethe Brueckner Goldston (BBG) ``data''~\cite{pg,brock}, variational calculation of Akmal {\it et ~al}.\,\cite{akmal}, Dirac Brueckner Hartree Fock (DBHF)~\cite{li}, non relativistic Brueckner Hartree Fock (BHF) with AV14 potential plus 3BF~\cite{baldo}, or heavy ion experimental data~\cite{daniel}, in the region where baryon density is around $2.0\rho_0-4.5\rho_0$, with $\rho_0$ the nuclear matter saturation density. This model also predicts a too large proton fraction ($Y_p$)~\cite{parada} which leads to a too low threshold density to start the direct URCA cooling process in neutron star. Furthermore, the predicted neutrino mean free path (NMFP) shows an anomalous behavior. On the other hand, there are parameter sets of RMF model which are specifically parameterized for interstellar purpose (e.g., parameter sets of Refs.\,\cite{pieka,glen}). This kind of parameter sets predicts a soft EOS at high densities but the predicted finite nuclei ground state properties are quite unsatisfactory. The possible reason is because the parameter sets with soft EOS have a relatively large nucleon effective mass ($M^*$) for finite nuclei density, which has a consequence that their spin-orbit splittings are relatively narrow compared with experimental data~\cite{anto}. This means that the  RMF model is not ``effective'' enough to cover a wide range of densities and the reason lies in the form of the nonlinear self coupling~\cite{brock}.

In 1996 the chiral effective Lagrangian model was proposed by Furnstahl, Serot and Tang~\cite{fst}, whose mean field treatment is hereafter called the ERMF model. This model has accurate predictions of the ground state properties of nuclei and the extrapolation to high densities yields a soft EOS which is consistent with other calculations~\cite{brock,akmal,li,baldo}, as well as experimental data of Refs.\,\cite{daniel,amu}. Nevertheless, it is widely known that this model still predicts a too large $Y_p$ in the neutron star~\cite{parada}. This is caused by the role of isovector terms, in which the corresponding parameters  are poorly constraint by insensitive isovector observables of finite nuclei.    

Therefore, in this report, we adjust the isovector-vector channel of the ERMF model in order to have a reasonable neutron star $Y_p$ without changing its ground state predictions in finite nuclei by fine tuning the symmetry energy ($E_{\rm sym}$) of the symmetric nuclear matter (SNM) to achieve a softer $E_{\rm sym}$ at high densities. We follow the prescription of Horowitz and Piekarewicz~\cite{pieka}  for the adjustment procedure. Besides that, we also study the anomalous behavior of the predicted NMFP in the neutron star. 

\section{Results}
In Table~\ref{CCG2} we list the parameter sets of the ERMF models (G2 and G2*) along with the parameter set of the standard RMF one. G2 is the standard ERMF parameter set, while G2* is a modified one with adjusted isovector-vector channel couplings.

\begin{table}[!b]
\tbl{
Coupling constants of the RMF models.}
{\footnotesize \begin{tabular}{@{}crcr@{}}
{} &{} &{} &{} \\[-3ex]
\hline\hline \\[-2ex] Parameter ~~ &~~G2~~ &~~ NL-Z~~&~~ G2*~~ ~~ \\[1ex]\hline
{} &{} &{} &{} \\[-1.5ex]
 ~~$m_S/M$ ~~       &~~0.554~~& ~~0.520~~&~~ 0.554~ ~~ \\[1ex]
 ~~$g_S/(4 \pi)$~~ &~~0.835~~& ~~0.801~~& ~~0.835~~ ~~\\[1ex]
 ~~$g_V/(4 \pi)$~~ &~~1.016~~& ~~1.028~~& ~~1.016~~ ~~\\[1ex]
 ~~$g_R/(4 \pi)$~~ &~~0.755~~& ~~0.771~~& ~~0.938~~ ~~\\[1ex]
 ~~$\kappa_3$    ~~ &~~3.247~~& ~~2.084~~& ~~3.247~~ ~~\\[1ex]
 ~~$\kappa_4$   ~~  &~~0.632~~& ~-8.804~~&~~0.632~~ ~~\\[1ex]
 ~~$\zeta_0$    ~~  &~~2.642~~& ~~0~~& ~~2.642~~  ~~\\[1ex]
 ~~$\eta_1$     ~~  &~~0.650~~& ~~0~~& ~~0.650~~ ~~\\[1ex]
 ~~$\eta_2$     ~~  &~~0.110~~& ~~0~~& ~~0.110~~ ~~\\[1ex]
 ~~$\eta_{\rho}$~~  &~~0.390~~& ~~0~~& ~~4.490~~ ~~\\[1ex]\hline
\end{tabular}
\label{CCG2}} 
\end{table}

\begin{figure}[!t]
\centerline{\epsfxsize=4.1in\epsfbox{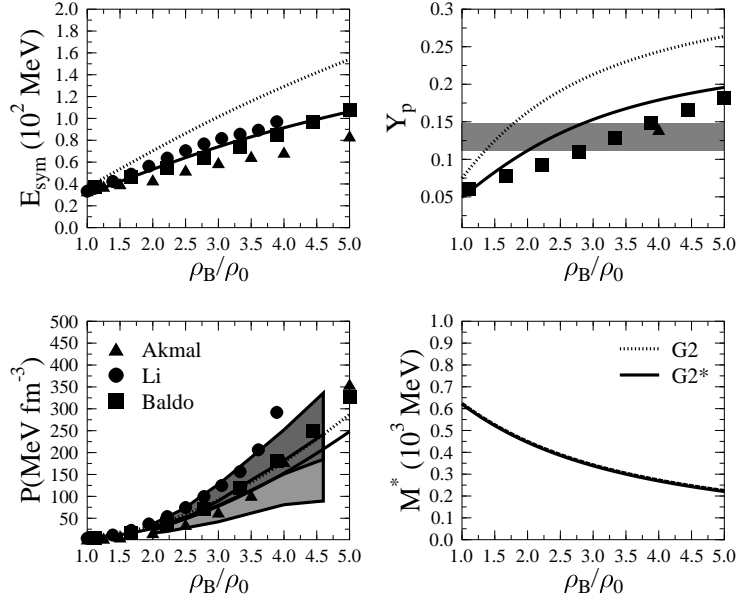}}   
\caption{EOS of the ERMF model at high densities. Shaded region in the upper-right  panel corresponds to the proton fraction threshold for the direct URCA process, while shaded region in the lower-left panel corresponds to experimental data. The results obtained by other calculations are also shown. \label{EOS}}
\end{figure}

The effects of the vector-isovector adjustment in $E_{\rm sym}$, $Y_p$, EOS and $M^*$ as a function of baryon density in the range of $(1-5)\rho_0$ are shown in Fig.~\ref{EOS}.  From the upper-left panel of Fig.~\ref{EOS} we can see that G2 and G2* have a similar value of  $E_{\rm sym}$ in the nuclear matter saturation density. Nevertheless, the G2* has a softer   $E_{\rm sym}$ compared to the G2 in higher densities. The agreement of the G2* $E_{\rm sym}$ with other calculations~\cite{akmal,li,baldo} also appears in this figure. 

Since we assume that the neutron star matter consists only of neutrons, protons, electrons, and muons, the relative fraction of each constituent can be  determined by the chemical potential equilibrium and the charge neutrality of the neutron star at zero temperature.  The results for $Y_p$ of both parameter sets can be seen in the upper-right panel of Fig.~\ref{EOS}. Since G2* has a softer   $E_{\rm sym}$ than G2, G2* has a higher threshold density ($\rho_B\approx 2.5 \rho_0$) for starting the direct URCA compared to the G2.  The predicted $Y_p$ of the G2* shows a better agreement with calculations of Refs.~\cite{akmal,baldo} rather than that of the G2.

From the lower-left and lower-right panels of Fig.~\ref{EOS}, it is obvious that both parameter sets have a similar trend in the pure neutron matter EOS and a same $M^*$, i.e., both have a soft EOS and have $M^* \sim 0.6 M$ at the nuclear saturation density, but large $M^*$ at high densities.  Although less significant, the G2* parameter set has a softer equation of state compared with the G2 one at higher densities. This fact indicates that quantitatively the neutron star properties (masses, radii, etc) predicted by both parameter sets are not too different.

\begin{figure}  
\centerline{\epsfxsize=2.5in\epsfbox{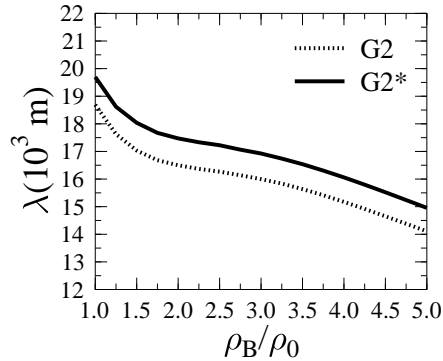}}   
\caption{Neutrino mean free path predicted by ERMF models.  \label{NMFP}}
\end{figure}

The results for the NMFP of both parameter sets are shown in Fig.~\ref{NMFP}. Clearly, the NMFP trend of both parameter sets is similar. No anomaly in the NMFP is observed, since both parameter sets have a large $M^*$ at higher densities, only their magnitudes are different. This means that the vector-isovector contributions have almost no influence in controlling the appearance of the anomalous behavior of the NMFP in the neutron star.

\section{Conclusion}

The standard RMF model needs additional nonlinear terms in order to simultaneously produce accurate ground state predictions of finite nuclei and realistic description in the high densities regime. The ERMF model has a strong theoretical ground to meet this requirement through the language of effective field theory. The reason that the ERMF model can cover a wide range of densities originates from the fact that the model has $M^* \sim 0.6 M$ at the nuclear saturation density, but a large $M^*$  and a soft $E_{\rm sym}$ at high densities. A more detailed analysis can be found in Ref.\,\cite{iso_prc}.

\end{document}